\documentclass[runningheads]{llncs}
\usepackage{graphicx}
\usepackage{todonotes}
\usepackage[hidelinks]{hyperref}
\usepackage{ulem}

\graphicspath{{images/}}
\begin{document}
\title{Smart technology in the classroom: a systematic review\\ {\small Prospects for algorithmic accountability}}
%
%
\author{Arian Garshi  \and
Malin Wist Jakobsen  \and
J{\o}rgen Nyborg-Christensen  \and
Daniel Ostnes  \and
Maria Ovchinnikova
}
\institute{Faculty of Social Science \\ Department of Information Science and Media Studies \\ University of Bergen}

\authorrunning{Garshi et al.}
\titlerunning{Smart technology in the classroom}
%
%
\maketitle              
\begin{abstract}
Artificial intelligence (AI) algorithms have emerged in the educational domain as a tool to make learning more efficient. Different applications for mastering particular skills, learning new languages, and tracking their progress are used by children. What is the impact on children from using this smart technology? We conducted a systematic review to understand the state of the art. We explored the literature in several sub-disciplines: wearables, child psychology, AI and education,  school surveillance, and  accountability. Our review identified the need for more research for each established topic. We managed to find both positive and negative effects of using wearables, but cannot conclude if smart technology use leads to lowering the young children's performance. Based on our insights we propose a framework to effectively identify accountability for smart technology in education.

\end{abstract}

\section{Introduction}

Today children are exposed to more technology than at any other point of history \cite{clements1998young}. This applies to the use of technology for both entertainment and education. Smart technology has mostly been evaluated in terms of how efficient it makes learning, how easy it is to use,  and how entertaining the applications are. However, as the use of smart technology is becoming widespread, there is also a need to study the impacts that these technologies may have on children.

Under smart technology in education, we consider learning applications, software, and tutoring systems as well as different types of wearables that can track student’s activity. How ethical is it to use technology with undocumented impact for children while they are in a class? Is this technology necessary? Is its efficiency for the learning process sufficient justification for its use? 


To understand the state of the art in studies of the impact of AI technology on education, specifically the impact of the wearables on young students' learning,  we conduct a systematic literature  review. We use the insights from the review to understand and outline the perspective for algorithmic accountability in the classroom technology context.  

Articles were identified and collected from Web of Science\footnote{\url{https://apps.webofknowledge.com/}}, Google Scholar\footnote{\url{https://scholar.google.com/}}, and ACM FaccT Conferences\footnote{\url{https://facctconference.org/}}, and the systematic review spans five topics: we aim to understand how young students are affected by surveillance, what technologies exist, why they are used, and how they are used by schools and educators. As the impact of wearables is studied by different research fields we wanted to ensure that literature in all the possibly relevant disciplines are explored. Our systematic review methodology is based on: a) proper definition of search strings, b) article relevancy, c) topic relevancy, d) successfully produce results for the research questions.

The literature review has shown that the majority of work on the use of AI and wearables in education mainly focuses on the innovative uses for  revolutionizing learning. Comparatively little attention is paid to the possible negative side-effects of constant monitoring.
Major areas of concerns in the literature were found to be breach of privacy, surveillance culture and the neglect of self-monitoring practices.

The systematic review has also shown relatively little work in accountability particularly in the domain of smart technology use in education. Young children at school are a vulnerable group and depend on guardianship or authorities to decide for them. Parents and teachers cannot be expected to understand technical details of all of the smart technology that can be used in the classroom. It is therefore important to properly establish rules of accountability in the use of wearable technology in education. The consequences of the use of wearables in the classroom are important to identify, and this includes identification of accountable parties to mitigate harm. We used insights from the reviewed work to advance further the discussion on accountability. Specifically we created an accountability framework to help identify the accountable parties for wearable technology in the classroom and ensure proper transparency in the lifetime of technology.

This report is structured as follows: We introduce our methodology in Section~\ref{sec:methodology}, showing our query designs and discussing the limitations of our review. Section~\ref{sec:aiineducation} covers the topic of AI technologies in education, what advantages and disadvantages it brings into the learning process. Section~\ref{sec:survinschool} presents the concept of surveillance in school. Section~\ref{sec:wearinschool} presents the literature on the use of wearables in schools. We then look at the psychological effects of being under surveillance, and what issues this might cause in Section~\ref{sec:psychinschool}. Section~\ref{sec:accountability} covers the topic of accountability in AI. Lastly we draw our conclusions and outline directions for future work in Section~\ref{sec:conclusion}.

\section{Methodology}\label{sec:methodology}
The systematic review was divided into five topics: AI in education, surveillance in school, wearable technology in school, psychological effects of surveillance, and accountability in AI. We discuss our approach to finding the relevant publications and its limitations.
\subsection{Query design}
For each of our topics of interest, we designed string queries to produce relevant results. Journal pages and databases that were queried include: Google Scholar, Web of Science  and ACM FAccT Conference Papers (2018, 2019 and 2020). Google Scholar effectively produces results that can be sorted by citation index and publication years. Web of Science is common for scientific and technical articles, hence its inclusion. The ACM FAccT Conference is a cross-disciplinary conference that focus on fairness, accountability and transparency in socio-technical systems and is thus included for its relevancy to work on accountability.

Our queries have the following structure
\begin{center}
    [wearables AND school]
    ["place in schools" wearable technology]
    [wearables AND (school OR education OR children)]
    [(accountability AND (AI OR artificial intelligence OR autonomous systems))]
    [(algorithm OR algorithmic OR algorithms) AND (accountability OR accountable OR accountabilities]
    ["effects of surveillance" AND "school" AND "learning" AND "children"]
    [(surveillance AND school (school OR education OR students OR student))]
    [((camera OR CCTV OR fingerprint OR "facial recognition" OR "metal detector") AND (monitor OR monitoring OR surveillance OR spy OR spying) AND (school OR kindergarten OR university OR college OR education OR student))]
    [Biometrics school]
    [Biometrics]
    [Social issues wearables]
    [Issues wearables]
    [cctv school]
    [("AI" OR "artificial intelligence") AND "education" AND "learning"]
    [Intelligent tutoring systems]
\end{center}

 Querying through Google Scholar produced results in English, and were sorted by relevance and high citations, but no specific year limit was set. As this is a wide project of multiple fields and mature topics, it is important not to exclude older work that either defined key aspects, or newer work that pushed the state of the art. Google Scholar includes books, direct citations and other reports, hence the queries produce many more results. Some older citations may also be excluded due to lack of tagging to match the produced queries. Web of Science produce lots of content for social sciences and computer science and also calculate an impact factor. The results of WoS are not as vast compared to Google Scholar, thus well-defined queries produces accurate results that were effectively sorted. The ACM FAccT Conference is a computer science conference with focus on Fairness, accountability and transparency in socio-technical systems. All papers produced in the years 2018, 2019 and 2020 have been screened from and identified.

\begin{figure}[!htb]
    \centering
    \includegraphics[width=12cm]{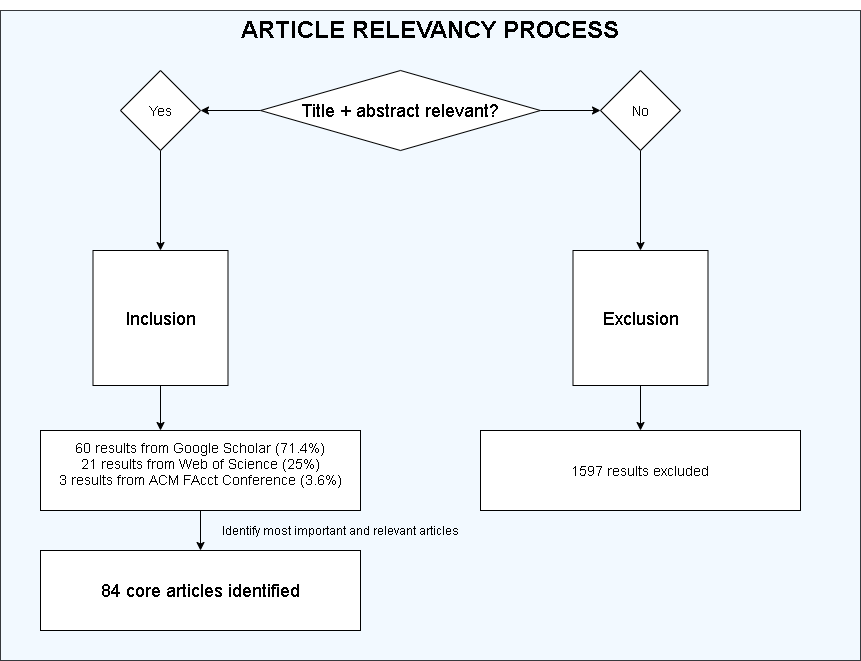}
    \caption{Flowchart of article selection process}
    \label{fig:fig1}
\end{figure}

 The articles we found were assessed for their relevance through screening  title and abstract. We considered a total of 1581 articles. Screening resulted with 84 core articles. We identified additional 15 articles following references. Exclusion of articles occurred either through the preliminary process of identifying core articles, or full reading of the article proved to be unfit for this projects scope. We reviewed 99 articles in detail. Figure~\ref{fig:fig1} visualize the article relevancy process and feature the article distribution from where they were found.

\subsection{Query Limitations}
Our hypothesis is that the use of wearables correlates with lower school performance for young children. To confirm this hypothesis, we explored research on children and their behavior when they are being observed in a classroom environment. As this is an emerging topic, research targeting our specific goals are limited. For this reason, we also explored literature that intersects with our topic. We looked at keywords such as "AI", "Surveillance", "Wearables", "Classroom", "Accountability", "Behavior", "Consequences", etc. We acknowledge that there might be some literature that we missed due to the restriction on queried keywords in addition to the volume of research papers that we were able to review.

\section{AI in education}\label{sec:aiineducation}
There are various ways to use AI in learning: students use virtual classrooms and gain new skills through various applications and educational online games. AI modules and Educational Data Mining techniques are able to track students' performance in order to provide a better learning experience based on students' needs
\cite{Baker2014EducationalAnalytics}. The development of such tools and technologies as teaching robots, intelligent tutoring systems (ITSs), and adaptive learning systems significantly change the learning process \cite{Chassignol2018ArtificialOverview}.

We discuss the impact AI technologies have on learning. First, we look at what advantages AI technologies bring into education. Second, possible issues of the use of AI in learning and teaching are going to be discussed.
 
\subsection{How does education benefit from using AI?}

The systematic literature review identified three different AI technologies that are being often used in learning: Intelligent Tutoring Systems (ITSs), intelligent support for collaborative learning, and intelligent virtual reality \cite{Luckin2016}. The benefits that these technologies bring are also discussed. 

\paragraph{\textbf{Intelligent tutoring systems}} 
Nwana \cite{Nwana1990IntelligentOverview} defines the Intelligent Tutoring System (ITS) as a computer program that is designed to incorporate techniques from AI to provide tutors that know what they teach, who they teach and how to teach. ITSs are able to perform one or more tutoring functions such as asking questions, assigning tasks, offering hints, providing feedback, or answering questions. These systems also use student input to model the student's cognitive, motivational or emotional states in a multidimensional space \cite{Nesbit2014HowEducation,Murray2003AnArt}. The researches on ITS have successfully delivered techniques and systems that provide personalized support for problem-solving activities in a variety of domains (e.g., programming,physics, algebra, geometry, SQL) \cite{ConatiAIModelling}.

Intelligent tutoring systems are able to offer considerable flexibility in the presentation of material and a greater ability to respond to idiosyncratic student needs. In addition, ITSs have been shown to be highly effective at increasing students' performance and motivation \cite{Beck1996ApplicationsEducation,PaviottiIntelligentOverview}.

\paragraph{\textbf{Intelligent support for collaborative learning}}
The concept of collaborative learning refers to “an instruction method in which students at various performance levels work together in small groups toward a common goal”. Thus, the success of one student helps other students to be successful since students are responsible for one another's learning as well as their own \cite{Gokhale1995CollaborativeThinking}.

In order to increase the level of the success of the group as well as learning, some approaches based on AI technologies can be useful: for example virtual agents and adaptive group formation \cite{Luckin2016}. Virtual agents can play the role of a tutor or a virtual peer which, according to Baylor, increases the motivation of the learners \cite{Baylor2009PromotingAppearance}. As for adaptive group formation, this approach helps to form a group best suited for a particular collaborative task using information about participants, for example, students’ cognitive level \cite{MuehlenbrockFormationInformation}.

\paragraph{\textbf{Virtual reality}}
Virtual reality (VR) can be seen as a form of human-computer interface characterized by an environmental simulation controlled only in part by the user. VR requires hardware and software that furnish a sense of immersion, navigation, and manipulation \cite{HelselVirtualEducation}. The main purpose of the use of VR in education is to explore and train practical skills, technical skills, operations, maintenance, and academic concerns \cite{Psotka1995ImmersiveTraining}.

Pantelidis \cite{Pantelidis1996VirtualEducation,Pantelidis2017ReasonsReality} claims that VR grabs and holds the attention of students. They find VR technology exciting and challenging to walk through an environment in three dimensions and interact with it. In addition, VR can more accurately illustrate some features and processes as well as change the way a learner interacts with the subject matter. Using VR in learning  encourages active participation rather than passivity.  

The study \cite{HusseinCarlNatterdal2015TheStudy} shows that there are a variety of possibilities to use VR in education for different purposes. In medicine, VR can allow users to perform tasks that carry safety concerns or cannot be achieved in real life, hence people are able to practice more. In fields of architecture and design, VR technology encourages users to be creative. The use of VR is also able to create interactive environments to teach kids about basic science facts and small lab simulations. Thus, virtual reality provides more opportunities for practice and training.

\subsection{Issues}

Despite the fact that applications and learning concepts that are based on AI technologies can have a significant positive effect on education, the use of AI in education brings  issues for teachers, students, and researchers \cite{Pedro2019ArtificialDevelopment}.

One of the issues mentioned is the need for a comprehensive public policy on AI development. Such public policy will give an opportunity to spread AI research. Today most of the AI products in education come from the private sector. If there is no partnership between state and private companies, public policy will not be able to cope with the speed of innovation. There is a need for such a partnership in order to be able to enhance AI training and research \cite{Pedro2019ArtificialDevelopment}.

There is also a concern that AI\&Education research does not provide an educational clear pedagogical alternative, besides individual tutoring. There is a specific view of the role of the computer in education: it is considered as a cognitive tool, which performs lower-order tasks and allows the user to concentrate on the more important higher-order thinking. In that case, developing such tools does not really need AI technology to be involved \cite{Andriessen1999WhereAI}.

One of the biggest issues for AI in education is privacy and data collection. People are concerned about how and when their data is collected, and most important for what purposes. Data collection becomes even more complicated in the context of young learners, who, in legal terms, cannot yet provide express consent regarding the collection and use of their personal data \cite{Pedro2019ArtificialDevelopment}.

Another issue that AI brings into education is that teachers might not be ready for such technologies. For educators, the use of new applications and ITSs systems in their teaching programs requires a rapid revision of what is taught and how it is presented to take advantage of evolving knowledge in a field where technology changes every few years \cite{Kolodner2013AIEducation}.

\section{Surveillance in school}\label{sec:survinschool}
Surveillance refers to the continuous observation of a place, person, group or activity in order to gather information, influence and manage individuals. 
Surveillance in schools has previously been largely the task of the teachers, hall monitors as well as simple attendance lists and grading systems. Technologies such as closed circuit television (CCTV) cameras, webcams, automated fingerprint identification systems, facial recognition, and metal detectors have seen an exponential growth of use in schools \cite{Hope2015GovernmentalityDevices}. 

In 2014, the US market for surveillance cameras, access control equipment and notification systems for schools and colleges was roughly 768 million USD. A large contributor of justification for such spending is the fear of rapid school shootings. For instance, after the 2012 Sandy Hook Elementary School shooting, the schools of Virginia were awarded 6 million USD to pay for video monitoring systems, metal detectors and other security upgrades \cite{Casella2018SchoolOfferings}. Such intense surveillance is argued to create prison-like conditions, specially for students in low-income areas \cite{Nance2014SchoolAmendment}.

We are interested in which technologies exist and how they are currently used by schools and educators. Although the surveillance of students is an international matter, the assessment of such technologies are based on their use in the United States and the United Kingdom due to the abundance of gathered data and research in the regions.  

\subsection{Video Surveillance}
CCTV cameras collect images through cameras and transfer the footage to a monitoring device where they are available to be reviewed \cite{Hope2009CCTVControl}. They present one of the most common surveillance methods in the schools internationally. An estimated 85\% of secondary schools in the United Kingdom, and two thirds of high schools in the United States use some form of CCTV systems \cite{studiesTheSchool}. 

The motivation of implementing CCTV are primarily crime prevention and detection as well as to detect vandalism, bullying, smoking, monitoring staff performance and to prevent intrusions by strangers \cite{Taylor2010IPrivacy}. However, it is common for schools to be careless about the privacy of students and have been found to not be in compliance with the data protection act of 1998 \cite{Taylor2011UK1998}.

The rampant use of CCTV has led to students comparing their school to a prison \cite{McCahill2010TheMums} and has raised paranoia to a degree that some students speculated that there were cameras in the toilets, that some cameras were hidden and could record their voice \cite{McCahill2010TheMums,Birnhack2018CCTVConsciousness}.

\subsection{Internet Surveillance}
Ever since the Internet gained popularity in the 1990s, classrooms have been connected to the internet to provide students with universal access to networked communications technologies. However, this has changed the classroom exponentially with the increased level of surveillance \cite{Steeves2018DigitalClassroom}.

The motivation of implementing internet surveillance systems has been concerns about online pornography, chat rooms, hate engendering websites, websites that sought to encourage experimentation with drugs and bomb creation, copyright violations, cyberbullying, piracy and hacking \cite{Hope2002SchoolRisk,Ahrens2012SchoolsState}. 

A number of private companies track the students on their personal social media accounts \cite{Shade2016HonestlyMedia}. The monitoring systems flag concerning phrases through the use of machine learning and artificial intelligence and alert school officials informing about the incidents. 

Filtering software is also a tool that is used to block access to online content that is deemed to be harmful. The Queensland Government’s Department of Education used school filtering software to block various social media websites, arguing that they did not offer educational value \cite{Hope2018UnsocialUse}.

A frequently reported problem is the fact that school filters block access to sites that the students need to visit to complete in-class assignments, causing frustration and disengagement from the learning material \cite{Steeves2018DigitalClassroom,Hope2013TheOver-blocking}. 

Research shows that some students intentionally engage in punishable online behavior, which can be understood as an escape from tedious routine through publicly testing boundaries. This act is seen as an important part of identity formation \cite{Hope2007Riskmisuse}.

\subsection{Biometric Surveillance}
Biometrics is the science of establishing the identity of an individual based on the physical, chemical or behavioral attributes of the person \cite{Jain2007IntroductionBiometrics}. Biometric technologies are used for surveillance in schools in the form of automated fingerprint identification systems, palm vein scanners, iris scanning devices and facial recognition software \cite{Hope2016Biopower2.0}.

This field is expanding rapidly and schools represent a significant market that uses such technology. In 2011, 2000 secondary schools and 2000 primary schools in the United kingdom had adopted some form of biometric technology. Three years later in the same country, it was estimated that 1.28 million secondary schools were fingerprinted \cite{Gray2018BiometricsSchools}. The motivation for the use of fingerprint identification has been registration, library book lending, cashless catering systems and personal lockers \cite{Hope2016Biopower2.0}.

The use of fingerprinting for identification seems intuitive in the sense that the body parts in question are merely patterns in the skin and are probably not integral to people's sense of uniqueness. It may even be less threatening to privacy than other surveillance methods such as scanning systems that produce an image of the naked body or manual frisking searches. However, the routine use of our body as a password may add new dimensions to our experience of embodiment in ways unimaginable to present generations \cite{Bryce2010BiometricCurriculum}.

The use of fingerprinting technology was one of the first applications of machine pattern recognition and is well established with powerful recognition systems that can match fingerprints with millions of matches per second. However, this process is very complex and is not a fully solved problem. Besides the possibility to hack such systems, it also is possible to present fake fingers to the sensors and be able to steal other people's identity \cite{Maltoni2Recognition}.

Metal detectors are another popular form of biometric surveillance used in schools. Biometric survelliance is one of the ways that the American schools are trying to solve the problem of weapons in schools. Six percent of public schools in the country are conducting daily or random metal detector searches of students \cite{Gastic2011MetalSchool}. 
Metal detectors have been argued to represent a source of fear by reminding students that their peers may be carrying weapons. Using metal detectors alienates students and decreases the students confidence in the school \cite{Schreck2003SourcesSecurity}. This fear could lead to students taking protective measures to defend themselves and paradoxically cause more violence. 

\subsection{Issues} 

The research shows that the use of surveillance technologies is rapidly growing \cite{studiesTheSchool}. Schools use tragedies such as school shooting (in the United States), bullying and suicidal students as mascots to justify the implementation of increasingly invasive surveillance systems to track and identify students both during and after school \cite{Casella2018SchoolOfferings,Taylor2013SurveillanceSchools,Shade2016HonestlyMedia}. Private companies that sell surveillance technology are earning massive amounts of money to track student in social media with seemingly little regard for their privacy \cite{Taylor2010IPrivacy} while fingerprint scanners and CCTV cameras track every movement and action of the students while they're at school \cite{Gray2018BiometricsSchools}. The negative reaction about the prison-like state, alienation and paranoia that students have reported to feel during school \cite{Nance2014SchoolAmendment,McCahill2010TheMums,Schreck2003SourcesSecurity,Birnhack2018CCTVConsciousness} raises major concerns about how wearables which are arguably even more invasive than the technologies that are covered in this Section may affect the students trust in their schools and their sense of liberty.

\section{Wearables in school}\label{sec:wearinschool}
Wearable technology can be defined as \textit{items worn with acceptable function and aesthetic properties, consisting of a simple interface to perform set tasks to satisfy needs of a specific group} \cite{Wilson2018WearableFuture}. As wearable technologies create opportunities in the educational domain it is important to understand their affordances and issues. Technology is getting smaller and is being created in everything and everywhere, even in toys \cite{Anderson2017WearableLearnTechLib,Holloway2016TheToys}. With that in mind, we introduce the fourth main theme of our systematic review, accountability of wearables in school. We have gathered around 20 scientific articles surrounding the theme of wearables in education, from the web of science and google scholar. The articles we have collected are only dated 2014 and onward. This is so they are relevant and holds the technical state of the art. The articles will be categorized after common themes found in the articles. 

\subsection{Possibilities of wearables in an educational environment}
\paragraph{Learn from learning}
The gathered articles state that there are multiple possibilities regarding using wearables with children in an educational domain. An emerging theme found in 2 of the 20 articles was about how we can learn from \cite{Gersak2019UseMovement,Goh2019AClassroom}. As one can use wearables to gather information about how well children learn and how to utilize the technology for better learning environment. Geršak et al. explore the possibilities of wearables in a classroom environment through a case study of learning geometry using movement \cite{Gersak2019UseMovement}. Goh, Carroll and Gillies summarized potential uses for portable tech in a classroom to improve learning and teaching practice's \cite{Goh2019AClassroom}.

\paragraph{First person view and simulation}
Introducing new technologies in an educational domain creates new methods of learning. Virtual reality wearables can be used to achieve a first-person view of learning new skills \cite{Coffman2015GoogleLearning.}. Examples could be showing how to perform surgery or simulating riskier scenarios \cite{Bower2015WhatTechnologies}. Another way is to enhance the learning experience in general, for example with the use of google glass as Coffman and Klinger explored in their conference paper \cite{Coffman2015GoogleLearning.}.

\paragraph{Helping children with disabilities}
Borthwick explores the possibility of student engagement through wearables such as Fitbit, GoPro cameras, Google Glass and Oculus Rift\cite{Borthwick2015}. The author also presents how children and students with physical disabilities can benefit from this new technology. For example, children with visual impairments can use google glass to help children navigate when walking, or using GPS connected wearables so parents get notified when children are out of "bounds" from where they should be e.g. school. This possibility is also explored by Gilmore et al., where parents have used devices to monitor children within geo-technological fences \cite{Gilmore2019SecuringWearables}. Another possibility of using this GPS technology is explored in the article by Freeman et al., where the classroom is sensor tagged so visually impaired children achieve more independence with wearing smart watches \cite{Freeman2017AudibleIndependently}.

\paragraph{Health}
Wearing Fitbit or other training wearables can be a motivation for being healthy, by tracking activities and setting personal fitness goals \cite{Ortiz2016ActiviTeen:Setting}. In their feasability study, Muller et al. show that wearables can be used as a motivation for children staying healthy \cite{Muller2018FeasibilityStudy}. Goodyear et al. focus on young people's use of Fitbit \cite{Goodyear2019YoungResistance}. They discovered that daily steps and calorie burning target did not engage young peoples engagement more than a few weeks. The different results from Ortiz, Muller and Goodyear show us that more research is needed to conclude whether wearables can be used as a psychical education motivation or not \cite{Ortiz2016ActiviTeen:Setting,Muller2018FeasibilityStudy,Goodyear2019YoungResistance}.

\subsection{Issues}

As there are multiple opportunities using wearables as a tool in education, it is important to highlight the issues surrounding them.

\paragraph{Safety of the children.}  Borthwick et al. \cite{Borthwick2015} explore how using wearables can affect student safety, which can be applied to our research question about young children. Data gathered from wearable devices can make information about children too available compared children not using wearables, as the data usually contains sensitive information. Sensitive information like location, what they do, how long they are at certain locations. Another safety aspect to consider is children who use wearables due to health reasons,  for example, drug infusion pumps \cite{Mills2016WearingDevice}. Mills explores a case where a drug infusion pump was hacked and \textit{"it was possible to remotely change the amount of drug administered"} \cite{Mills2016WearingDevice}. This is not necessarily a wearable device such as a Fitbit, but as new technology becomes available it might be the next step when it comes to unsecured devices.

\paragraph{Distraction and dependence on technology}
Children using wearables in a classroom environment might become too dependent on this new technology, as children might not be able to apply the same knowledge for real-life scenarios when not using the wearables as in school\cite{Attallah2018WearableEducation}. Another issue with wearables is how they might be a distraction for other children in a classroom environment \cite{DeArriba-Perez2017TowardsEducation}.

\paragraph{Surveillance of Young children}
 Holloway and Green highlight the dangers with having technology in everything through sensors children are interacting with \cite{Holloway2016TheToys}. Showing the dangers on how something simple such as toys can have weaknesses that can be exploited. Another danger of using wearable technology with such young children is that the surveillance of children might become normalized. As parents are usually accepting of wearables as they are easy to use and the children like them \cite{Mackintosh2019ParentalStudy}. As Ching et al. explore the dangers on how wearable technology have authentication issues. Thus introducing surveillance dangers to young children \cite{Ching2016}. 

\section{Psychology of School Surveillance}\label{sec:psychinschool}
In order to determine if using wearable AI surveillance gadgets causes lower performance in young children, we have to look at the psychological effects of being under surveillance, and what issues this might cause. The literature review uncovered a worrying lack of research regarding the psychological effects of surveillance in the classroom, and what implications this have for student performance. AI powered surveillance gadgets and their effects on students as a psychological topic are suffering from serious neglect. However there are studies on motivation, creativity and the consequences of being watched that can shed some light of the possible psychological consequences of this new technology, and its application in the classroom. 

\subsection{Surveillance, Reward and Motivation}
One important aspect of how surveillance affects student performance, is to consider its effect on students motivation \cite{LepperTurningMotivation.,Deci1999AMotivation,Plant1985IntrinsicStyles,Enzle1993SurveillantMotivation,Hennessey2015IfClassroom}. These papers investigates the effects of surveillance on intrinsic and extrinsic rewards and motivation. An extrinsic reward is a tangible and visible reward given to an individual for achieving something. This is the opposite of an intrinsic reward, which is an intangible award of recognition, such as a sense of achievement or satisfaction from completing a task. Intrinsic motivation refers to behaviour that is driven by intrinsic rewards, and the motivation for this behaviour arises withing the individual because its naturally satisfying to them. This contrasts with extrinsic motivation, where one engages in a behaviour in order to gain external rewards or avoid punishment.

Lepper and Greene \cite{LepperTurningMotivation.} did a study of 80 preschool children (4-5 years old), where the children were to complete puzzles. Participants were divided into three conditions, the expected reward condition, the unexpected reward condition, and the no reward condition. There were three surveillance conditions. In the non-surveillance condition, the television camera had its lens removed and was faced away from the table were the subject was doing the puzzle, and there was no mention of the subject being under surveillance. In the low-surveillance conditions the light that indicated that a subject was under surveillance was turned on for one of the six puzzles. In the high-surveillance condition the light was turned on during four of the six puzzles. Three weeks after the individual experimental session was completed, the researchers measured the students intrinsic interest in solving puzzles, by having children choose between the target activity (solving puzzles) and a variety of other activities. Plant and Ryan \cite{Plant1985IntrinsicStyles} did the same experiment, and found that the same results applied to college students. 

Results \cite{LepperTurningMotivation.,Deci1999AMotivation,Plant1985IntrinsicStyles} show that the expectation and receiving of an extrinsic reward for engaging in an activity produced decreased intrinsic interest related to engaging in that activity. Surveillance produced an even greater additional decrease in later interest in an activity, and it also reduced participants autonomy \cite{LepperTurningMotivation.,Enzle1993SurveillantMotivation}. Autonomy refers to self-government over ones own actions, and the feeling of autonomy decreases under surveillance. Intrinsic motivation was greater in the no-surveillance conditions. One interesting finding from these experiments is that it made little difference whether the surveillance was constant or only occasional \cite{LepperTurningMotivation.,Enzle1993SurveillantMotivation}. It also interesting to note that if subjects were informed that the surveillant was watching because he was personally interested or curious as to how people would approach an activity it neither challenged personal autonomy nor undermined intrinsic motivation \cite{Enzle1993SurveillantMotivation}. 

\begin{quotation}
"The knowledge that
one's performance at a task is being observed
and evaluated by someone else, even when
there is no explicit expectation of any tangible
reward for engaging in the activity, appears
sufficient to decrease later interest in the task." \cite{LepperTurningMotivation.}.
\end{quotation}

Several studies has also looked at how intrinsic motivation is a precursor to creativity and that extrinsic motivation has a detrimental effect on it \cite{Amabile1979EffectsCreativity,Hennessey2003TheCreativity,Hennessey2015IfClassroom}. These studies suggest that surveillance is the killer of intrinsic motivation and creativity. 

\subsection{How Surveillance Influences Student Behaviour}
Several studies show how different aspects of surveillance can affect students behaviour online \cite{Dawson2006TheBehaviour,Hope2005PanopticismSchools}, self-monitoring \cite{Plant1985IntrinsicStyles}, cheating and pro-social behaviour \cite{Jansen2018TheBehavior}, moral judgements \cite{Bourrat2011SurveillanceCondemnation} and children's experience of trust, risk and responsibility \cite{Rooney2010TrustingResponsibility,KnoxAEnvironments}. 

Dawson \cite{Dawson2006TheBehaviour} investigates how online surveillance of students affect their behaviour. Through surveys Dawson \cite{Dawson2006TheBehaviour} studies the extent to which students perceived they changed their own online behaviour as a result of institutional surveillance techniques. The results showed that "all students indicated that browsing behaviour, range of topics and writing style is influenced by the various modes of surveillance" \cite{Dawson2006TheBehaviour}. This change in behaviour is explained through the fact that "when people are objectively self-aware—aware of themselves as an object or as viewed by another—they are likely to regulate themselves controllingly (i.e., as if they were concerned about  an other's evaluation of them) \cite{DeciTheBehavior.}. One interesting finding is that students unaware of the specific surveillance measures enacted by the institution performs a high level of self-regulation  \cite{Dawson2006TheBehaviour}.

Williamson \cite{Williamson2017} and Manolev et al. \cite{Manolev2018LearningCulture} investigates the ClassDojo application, which is a school based social media platform that incorporates gamified behaviour shaping functions. The app has been marketed as promoting positive psychological concepts such as growth mindsets and character development \cite{Williamson2017}. However the problem with this application is that it "requires teachers to monitor students constantly, catching students performing particular behaviours, generating, storing and analysing data through its software as this occurs" \cite{Manolev2018LearningCulture}.

\subsection{Issues}
\paragraph{Erosion of enjoyment} AI wearables can be seen as a form of extrinsic motivator resulting in either an extrinsic reward or punishment for the student. When students are under constant evaluation, the motivation for doing any task is inherently extrinsic, and the students might be awarded with praise or punished for bad results. One negative effect of extrinsic motivation is that it can lead to the erosion of enjoyment in tasks that were previously extrinsic. The results from Results \cite{LepperTurningMotivation.} clearly shows how the students enjoyment of playing with puzzles eroded, after being in the surveillance group for this study. 

\paragraph{Autonomy}The real danger of widespread adoption of wearable AI surveillance gadgets, from a psychological perspective, is that they will undermine a whole generations intrinsic motivation to learn. This will also be detrimental to children's autonomy, as the constant surveillance will influence how they interact with the world around them, as shown by Dawson \cite{Dawson2006TheBehaviour}. Dawson \cite{Dawson2006TheBehaviour} showed that when students are under surveillance they restrict their actions which negatively impacts their autonomy. For people to feel autonomous they need to feel that they are in control of their own lives, this control is reduced when they are under surveillance. 

\section{Accountability}\label{sec:accountability}
\subsection{Literature review}
Algorithmic accountability is a multidisciplinary theoretic field \cite{Wieringa2020} and examines an algorithm's influence, mistakes or biases \cite{Diakopoulos2016}. Algorithms are defined as instructions fed to a computer \cite{goffey2008algorithm,Wieringa2020} and in this systematic review, algorithms or algorithmic systems refer to AI based technology that is used in an educational environment by young students. Issues may arise with the use of AI in education \cite{Pedro2019ArtificialDevelopment,Kolodner2013AIEducation,Andriessen1999WhereAI,Borthwick2015,Mills2016WearingDevice,Attallah2018WearableEducation,DeArriba-Perez2017TowardsEducation,Holloway2016TheToys,Mackintosh2019ParentalStudy,Ching2016} hence the importance of properly defining accountability and identifying stakeholders in an artificial technology devices' lifetime. Accountability can be defined as
\begin{quotation}
 a relationship between an actor and a forum, in which the actor has an obligation to explain and to justify his or her conduct, the forum can pose questions and pass judgement, and the actor may face consequences \cite{Bovens2007}. 
\end{quotation}

One can argue that companies must assume ethical responsibilities for their creations instead of outsourcing the ethical concerns to those directly related to the implementation of the technology. So, who is accountable for the use of technology? This is not an easy question to answers, and researchers argue whether designers, users, stakeholders, or all of the mentioned are the ones accountable \cite{Dignum2017,Orr2020,Kroll2017,Martin2019,Kraemer2011,CerrilloIMartinez2019}. By defining accountability and describing factors related to it, we hope to find ways to describe responsible parties and create more transparency in the use of artificial intelligence in education. 

Decision-making algorithms must represent moral values and societal norms related to its operational context in order to ensure accountability. Requirements for accountability in artificial intelligence are guiding of actions (making decisions and forming belief), and explanation (assigning decisions to a broader context and classifying them onto moral values) \cite{Dignum2017}. Dignum \cite{Dignum2017} investigates how researchers should approach AI system design and how complex that can be. Current machine learning and deep learning technologies are not able to link decisions to input in meaningful ways, which means we are not able to understand their choices. Problems such as who is to blame when a self-driving car runs over someone must be in place in a responsible AI system, and mention how participation is essential; one must understand different cultures and lifestyles, therefore a framework that is effective across cultures must be put in place. The author propose the ART principles: accountability, responsibility, transparency. Investigations into each of these principles are essential to develop a responsible AI system \cite{Dignum2017}. 

Orr and Davis \cite{Orr2020} provide analysis of ethical responsibility in artificial intelligence through interviews of field experts. The problems of ethics in AI are more clear, than the ethical responsibility. Interviewees revealed that practitioners involved in design and development are the obvious choice as the accountable party as they are the creators of the product. However, the subjects argue the designers are but a single node in a more complex system. In regards to accountability norms, the interviews reveal a relationship between the nodes where one has power (legislators, organizations, clients) and the others have technical expertise (practitioners). Hence, it is not possible for practitioners to operate in full discretion but they must display independent results. Three stages are identified as nodes in the system, and put together they represent the process of defining, development, and deployment:  
\begin{enumerate}
    \item Organizations define parameters, 
    \item practitioners develop hardware and software, 
    \item deployment to users and machines.
\end{enumerate}

Results \cite{CerrilloIMartinez2019,Diakopoulos2016,Dignum2017,Garfinkel2017,McGregor2019,Orr2020,Shah2018} indicate the importance of understanding ethics as a collaborative process that is developed through practices and negotiations for the future of AI and the importance of accountability norms in case systems go wrong. Decisions must be clear and made under a set of rules that define legal fairness and transparency, identifying accountable collaboration between computer science, law and relevant fields \cite{Kroll2017}. Torresen \cite{Torresen2018} presents the importance of control mechanisms in systems to ensure collaboration between stakeholders, designers, and users \cite{Orr2020}. Hence, there is need for a regulatory institution to oversee proper implementation \cite{Koene2019}. The ACM Europe Council Party Committee (EUACM\footnote{\url{https://acm.org/euacm}}) and ACM U.S. Public Policy Council (USACM\footnote{\url{https://usacm.org}}) have together described and codified a set of principles to ensure fair use in technological systems: Awareness, access and redress, accountability, explanation, data provenance, audit-ability, and validation and testing \cite{Garfinkel2017}. 
\begin{quotation}
 "Accountability rejects the common deflection of blame to an automated system by ensuring those who deploy an algorithm cannot eschew responsibility for its actions."\cite{Garfinkel2017}
\end{quotation}
Martin \cite{Martin2019} argues that firms who develop algorithms are responsible for the ethical outcomes of their technologies. If some algorithm acts  to influence people, then the companies should be accountable for the influence of their technology. By visualizing the delegation of responsibilities in algorithmic design, the author conclude that the creators of the algorithm decides the structure and allocation and should thus be held accountable. Derived from this logic, inscrutable algorithms constructed as difficult to understand and hard to explain, will prove greater accountability for the designer as it excludes the involvement of users in the decision role \cite{Martin2019}. This is defined as the {\it accountability gap} \cite{Schultz}. If an algorithm does impose ethical concerns, they either implicitly or explicitly take a stand on an ethical issue. Designers should therefore leave ethical issues for the users or create algorithms that are transparent and easy to understand \cite{Kraemer2011}. Although algorithms are more effective and mostly time-saving, they can make administrative powers less transparent and accountable for their decisions. Public administrations must warrant their use by knowing what data they use, what results are produced and what results they were expecting \cite{CerrilloIMartinez2019}. It is important to expose the design decisions and nature of algorithms in systems that consist of interconnected algorithms to identify accountability and liability. Multiple initiatives can help improve this: 
\begin{enumerate}
    \item  Pilot test the model to investigate bias and include stakeholders to have clear definition of what you are trying to achieve
    \item  Publish the model and the data to explain its provenance
    \item Monitor outcomes for differential impacts (e.g., focus on minorities
    \item   Right to challenge and redress (e.g., EU's General Data Protection Regulation
    \item   Better governance to improve oversight
    \item  The private sector should adhere to the standards of accountability and governance
\end{enumerate}
 
Diakopoulos \cite{Diakopoulos2016} discusses the importance of understanding the types of decisions an algorithm can make: prioritize, classify, associate and filter. Governments provides social goods and uses its power through moderated norms and regulations and is accountable for the citizens. On the other hand, private organizations do not have the same public accountability regulations. Robust policy and a transparency standard could solve this organizational gap. Diakopoulos \cite{Diakopoulos2016} proposes five categories to consider disclosing: 1) Human involvement: explain the goal, purpose and intent of the algorithm 2) Data: Share data accuracy, completeness, timeliness, uncertainties and other limitations 3) The model: Share the model and the modelling process 4) Inferencing: Benchmarking and testing to identify further limitations Algorithmic presence: Inform when the algorithm is in use or not \cite{Diakopoulos2016}. McGregor et al. \cite{McGregor2019} apply international human rights law (IHRL) as a framework for algorithmic accountability, and identify five factors for effective accountability: 1) A clear understanding and definition of "harm" 2) Full overview of the design, development and deployment cycle of an algorithm 3) Nodes of power \cite{Orr2020} must have clear obligations and responsibilities 4) Clear guideline for emergency management of harm caused 5) Focus on accountability measures throughout the life cycle of an algorithm \cite{McGregor2019}.  

Algorithmic decisions are not objective, but are designed for a set of actions \cite{Martin2019a}. Organizations want to conduct good business, but mistakes are unavoidable. Mistakes can come through bad inputs, bad reasoning and bad executions. In order to produce good decisions, organizations apply algorithms and autonomous systems to help this process. However, the algorithms also produce biased decisions and make mistakes. It is therefore important to identify mistakes in algorithmic decision making, and these fall into two classes 1) Category mistakes and 2) Process mistakes \cite{Martin2019a}. Category mistakes are algorithms that categorize individuals and are prone to two mistakes which are false positives (labelled as paying attention when they are not) and false negatives (not labelled as paying attention when they are). Process mistakes occur when an algorithm makes a fault in how a decision is made, disregarding the outcome.

As algorithms often "learn" factors that are important from current data, they may use unsuitable factors even when designed not to do so \cite{Martin2019a}. Designing accountability for mistakes produces better algorithmic decisions as they can identify responsibilities related to handling of mistakes \cite{Kanellopoulos2018,Martin2019a,Orr2020}. Koene et al \cite{Koene2019} study governance frameworks for algorithmic accountability and transparency. Algorithmic Impact Assessment (AIA) is a framework to support policymakers to understand an algorithmic system, where they are used, and allow the community to raise mitigation issues \cite{Koene2019}.
\subsection{Framework}
Sandall writes" \textit{``Wearable technology has the potential to impact schools in the same way as the computers and mobile devices of today''} \cite{Sandall2016WearableHere}. With this in mind, it is important to create boundaries and identify how smart technology should be implemented, as mentioned by Strandell \cite{Strandell2014MobileControl}.

We identify the need for a regulatory framework to govern algorithmic accountability with wearables in school. Based on insights from our literature review, we propose the following framework to identify accountable parties:
\begin{enumerate}
    \item The establishing of a regulatory institute
    \item Publication of proposed technology.
    \item Publication of scope, intended use and the technology's effect on the educational process
    \item Properly test and identify potential faults, bias and harm.
    \item Properly identify and notify necessary stakeholders. Technology must be transparent and explainable to non-technical stakeholders.
\end{enumerate}
The technology must be made publicly available along with its intended use, scope and effects on education (e.g. open source). Faults, biases and risk management must be identified and tested by professionals (e.g. engineers, artificial intelligence experts, technology designers, educators, psychologists). Open source would produce transparency, hence the need for a regulatory institute where experts can inspect the technology and make it explainable to the non-technical stakeholders. Further, stakeholders can issue complaints and accurately be provided with explainable action and identification of accountable parties.

For a technology to be adequate, there must be: 1) no risk of biased decision-making  to impact a students progress 2) no faulty technology to physically endanger the user, provoke fatigue, affect concentration or be uncomfortable 3) no possibility of the technology being hacked or misused 4) considerations for privacy 5) clear guidelines for; when to use it, how to use it, emergency management.

The information on the scope must be communicated through relevant venues, fora, and other platforms that include stakeholders (e.g. websites, email, conferences). Inclusion of all stakeholders is essential to manage the accountability gap. Stakeholders include: 1) Creators of the technological idea (organization, government) 2) Creators of the technology (engineers, programmers, designers) 3) Users of the technology (schools, teachers, students). In the case of young students, guardianship or authorities must be included as they make decisions while the students are below legal age. The stakeholders must accept the usage of the proposed technology either through communication with the school or regulatory institute, and should be made part of an obligatory assessment before a child enrolls into a school.

The regulatory institute will vary based on cultural differences (region, country, municipality) but must include the same core components: 1) Be an agent where the public (e.g., parents, teachers, school staff, students) can voice their concern. 2) Actively participate in the design, development and deployment phases of proposed technology. 3) Must approve the use of proposed technology. 4) The institute must include experts from multiple fields, such as: wearable technology, psychology, education, artificial intelligence, engineering and law. We consider the regulatory institute as a necessity for the development of smart technology because it closes the accountability gap by identifying accountable parties in the life cycle of smart technology. Further addenda on the framework can include research into law and the professional conduct of technology developers.

\section{Discussion}

Table~\ref{Tab:tab1} summarizes all the affordances  and issues in regard to four of the five topics of the impact of AI on education that we investigated. The topic of AI and accountability will be discussed separately.

\begin{table}[!htb]
\resizebox{\textwidth}{!}{%
\begin{tabular}{l|l|l}
\hline
 & AI in education & Surveillance in schools \\ \hline
Affordances & \begin{tabular}[c]{@{}l@{}}- Personalized support \\ - More flexible presentation material \\ - Better respond to needs\\ - Increasing performance and motivation\\ - Virtual agents increase motivation \\ - More training and practices \\ - Increase of attention\end{tabular} & \begin{tabular}[c]{@{}l@{}}-    Prevention of crime \\ -    Prevention of vandalism \\ -    Prevention of bullying\\ -    Detection of suicidal behavior\\ -    Blocking piracy\\ -    Blocking pornography\end{tabular} \\
 &  &  \\
Issues & \begin{tabular}[c]{@{}l@{}}- Privacy \\ - Rapid revision of learning plan\\ - Does not provide a clear pedagogical alternative\\ - AI technologies provided by private sector\end{tabular} & \begin{tabular}[c]{@{}l@{}}-    Breach of privacy\\ -    Prison-like state\\ -    Alienation\\ -    Paranoia\\ -    Unintentional blocking\\ -    Prone to hacking\\ -    Decreased trust\\ -    Increased protective measures by students\end{tabular} \\
 &  &  \\ \hline
 & Wearables in education & Psychological effect of surveillance \\ \hline
Affordances & \begin{tabular}[c]{@{}l@{}}- Learn from learning\\ - First person view \\ - Simulation of situations\\ - Helping children with disabilities\\ - Health\end{tabular} & - Wearable tech less obtrusive \\
 &  &  \\
Issues & \begin{tabular}[c]{@{}l@{}}- Safety of children\\ - Distraction\\ - Dependent\\ - Surveillance\\  \end{tabular} & \begin{tabular}[c]{@{}l@{}}- Decreased Interest\\ - Erosion of enjoyment\\ - Undermine intrinsic motivation to learn\\ - Self monitoring\\ - Lack of autonomy\\ 
\end{tabular}
\end{tabular}
}
\caption{Thematic overview}
\label{Tab:tab1}
\end{table}

\vspace{5mm}

\paragraph{}Table~\ref{Tab:tab1} shows that AI technologies bring flexibility in the presentation of material, personalization, and increase motivation. We have listed issues that are more general regarding the use of AI in education. We did not find much research done on the negative influence of AI technologies on students’ performance particularly. There can be several reasons for it. It is possible that it is complicated to get consent from the parents to be able to process and analyze the students’ personal data. It also can be connected to the fact that using AI applications in schools is somehow new phenomena as well as can be expensive to have. In addition, the fact that AI is popular and used in various aspects of our lives make people focus on its positive aspects more, but we still need to understand that the negative effects and problems with AI should also be highlighted in detail.


Although the affordances for the various surveillance methods that were reviewed are noble, a number of concerning reactions are revealed by the students. Hence, it is to be argued that this prison-like state that rampant surveillance is presenting at schools, breaches trust for all parties. Teachers lose their trust for the students ability to self-monitor, while the students lose their trust in their school's good faith as they are treated as suspects. We believe that the goal of schools should be to provide a safe and encouraging environment for the students to learn and discover themselves, and the heavy use of surveillance clouds this goal by treating students as potential wrongdoers. 

Schools are supposed to teach children how to become functional and contributing members of society as they become adults, and prepare them to handle themselves independently. As the use of wearable technologies is emerging and there is no long-term studies done on this phenomenon. It leaves us to argue that it might be counter productive to micromanage every single move that the students make to make sure that they fall in line without being able to evaluate themselves. 

Wearables in education show great promise on the amount of affordances, see Table~\ref{Tab:tab1}. Ranging from helping children with disabilities to health benefits. There was a lot of researchers that utilized wearables as a data collection tool in the classroom, but not researching on the consequence of use over time. The lack of research in this domain might be because of how new wearable technology is. There might be more of this research over time as wearables are getting cheaper and more accessible. If we look at the issues listed, one might say that the issues have larger consequences than the affordances, thus making the issue outweigh the affordances.

The literature in psychology indicates that the use of wearables correlates with lower school performance in young children. However, none of the psychological  studies we found used wearable technologies. We hypothesize that wearables might be a less intrusive surveillance technology, than the traditional surveillance cameras. We also hypothesize that because wearables have a more personal feel to them, some of the negative effects of surveillance might be avoided. More research is needed to be able to address these hypothesises. Due to the lack of literature it is hard to say whether or not wearable technologies have a direct negative impact on school performance in young children. 

We also found that the way the students were being told that they were under ``surveillance'' affected how they perceived this surveillance, and thus also their performance. This indicated that more work should be done with regards to how new technologies are introduced into a classroom setting, in order to reduce potential negative consequences. 

 We identify accountability as a process that includes multiple stakeholders and must thus be treated as such instead of as confined entities. The gap between government bodies and private organizations must be closed, and regulatory frameworks can prevent the renouncement of accountability in both sectors, hence our framework contribution. By strictly defining accountability relationships, we think that focus should shift onto design frameworks and methods  to create responsible and safe use of AI systems that can properly identify responsible agents. Through transparent design, development and deployment it will be easier to understand the use and implementation of wearables in school. We consider that accountability must at all times be identified, and the technology itself must be explained and tested to prove its use and disprove bias and potential faulty decision-making. Further development or changes must be explained and accepted before they are implemented. Only then can the technology be fully transparent and successfully delegate accountability. To close the accountability gap, we propose an accountability framework.

We describe some limitations to our research. Firstly, there are lack of studies dealing with psychological effects of new technologies in the classroom. Future work should examine how ``being under surveillance'' affects students, and how to present the fact that students are indeed under surveillance. Secondly, there is limited literature on educational surveillance that uses AI, which led to us researching general surveillance systems and drawing parallels as to how AI assisted systems may impact schools. Thirdly, the amount of articles and research scope must also be considered as a limitation factor. As we did not have access to technical product details and their applications, we had to investigate general concepts.

\section{Conclusions} \label{sec:conclusion}
The purpose of this review was to investigate whether the use of wearables correlates with lower school performance in young children, and if it's possible to identify  accountable stakeholders for wearable  technology in the classroom. The research indicates that the surveillance aspect of using wearables in school can have a negative impact on student performance. However, the extent of this impact is not entirely clear, and more research is needed, especially in regards to surveillance using AI.

In order to bring some understanding, research regarding different aspects of the use of AI at schools is needed: how can smart technology affect the performance of the students? What kind of data should be collected in order to present the best recommendation for a learning program for a particular student? How does the AI algorithm work? Can it fail at some point?

The systematic literature review identified that there is a need for more research for each established topic, particularly how and in which degree wearables can influence the performance. We managed to find some positive effects as well as negative, but we cannot conclude if using wearables with AI application can lead to lower students' performance directly. There are a lot of factors that have to be considered. Identification of accountable parties is a step in the right direction to produce transparent wearable technology that mitigate harmful consequences and create helpful utilities. Through our proposed framework it could be easier to implement wearable technologies in education and accurately identify affordances and issues. 

\bibliographystyle{splncs04}
\bibliography{references} 
\end{document}